\documentclass[9pt, a4paper]{article}   	
\usepackage{geometry}                		
\usepackage[]{graphicx}				
\usepackage{here}
\usepackage{caption}
\usepackage{subcaption}
\usepackage{amssymb}

\usepackage{authblk}
\title{Quantum Annealing Mechanism as A Measurement Process\footnote{This work is based on results from a project commissioned by the new Energy and Industrial Technology Development Organization (NEDO).}}

\author{K.Imafuku}
\affil{National Institute of Advanced Industrial Science and Technology}
\date{}
\begin{document}
\maketitle
\vspace{-1cm}
\begin{abstract}
An idea for an application of the quantum annealing mechanism to construct a projection measurement in a collective space is proposed. We use the annealing mechanism to drive the pointer degree of freedom associated with the measurement process. The parameters in its problem Hamiltonian is given not as classical variables but as quantum variables (states). By additionally introducing successive short interactions so that the back reaction to the quantum state (to be measured) can be controlled, we  invent a quantum mechanically parametrized quantum annealing process. Applying to a particular problem of discrimination of two collective states , we find that the process by the quantum mechanically parametrized annealing arrives at projection measurement in the collective space when the parametrizing quantum variables themselves are orthogonal (or distinguishable). 
\end{abstract}

Lots of attentions have been attracted to quantum annealing computations as an advanced computation technology \cite{PhysRevB.39.11828,PhysRevE.58.5355,2000quant.ph..1106F,2014McGeoch}. In the last decade,  not only theoretical aspects, but also implementations has been pursued harder than ever. Besides computations, it must be exciting to think about applications of quantum annealing mechanism itself. To make toward this direction, we consider an application of the mechanism to a measurement process distinguishing two collective states, as the simplest trial. Let us consider a situation where we are given one of the two states:
\begin{equation}\label{collective states}
|\Phi^{(N)}\rangle=|\varphi\rangle^{\otimes N}~\left(\in {\mathcal H}^{\otimes N}\right),\quad\mbox{and}\quad
|\Psi^{(N)}\rangle=|\psi\rangle^{\otimes N}~\left(\in {\mathcal H}^{\otimes N}\right)
\end{equation}
where
\begin{equation}
|\varphi\rangle=\sqrt{\frac{1+\epsilon}{2}}|0\rangle+\sqrt{\frac{1-\epsilon}{2}}|1\rangle,\quad
\mbox{and}\quad
|\psi\rangle=\sqrt{\frac{1-\epsilon}{2}}|0\rangle+\sqrt{\frac{1+\epsilon}{2}}|1\rangle.
\end{equation}
Suppose that we guess the given state by performing some quantum measurements on it. In the following, we assume that $N$ is large enough to make the two states approximately distinguishable, that is 
\begin{equation}\label{f-condition}
O(N\epsilon^2) > 1. 
\end{equation}
Under this assumption, one possible way to guess the given state is to perform $N$ individual measurements. For instance, performing a measurement defined by a set of projection operators $\{|0\rangle\langle 0|,~|1\rangle\langle 1|\}$ on each subspace ${\mathcal H}$, one obtain the $N$-bit sequence of $0$ and $1$. With the sequence,  by statistically estimating bias on the appearance frequencies of $0$ and $1$, one is able to make a correct guess. Another way is to perform a projection measurement on the collective state that is defined by a set of projection operators
\begin{equation}\label{collective projection measurement}
\{\hat{\Pi}_{\Phi}^{(N)},\hat{\Pi}_{\Psi}^{(N)}\}
\end{equation}
on total Hilbert space $\in {\mathcal H}^{\otimes N}$ such as
\begin{equation}\label{property 1}
\langle\Phi^{(N)}|\hat{\Pi}_{\Phi}^{(N)}|\Phi^{(N)}\rangle=
\langle\Psi^{(N)}|\hat{\Pi}_{\Psi}^{(N)}|\Psi^{(N)}\rangle=1
\end{equation}
and
\begin{equation}\label{property 2}
\langle\Psi^{(N)}|\hat{\Pi}_{\Phi}^{(N)}|\Psi^{(N)}\rangle=
\langle\Phi^{(N)}|\hat{\Pi}_{\Psi}^{(N)}|\Phi^{(N)}\rangle=0
\end{equation}
hold. In this strategy, one obtains $1$-bit information that directly indicates the given state, instead of the $N$-bit sequence in the previous way. In this sense, the collective strategy can be more efficient than the individual one. As the consequence of von Neumann postulate \cite{von2018mathematical} about the state after projection measurements, the projection measurement in (\ref{collective projection measurement}) preserves the collective state as was originally given, i.e.,
\begin{equation}\label{property 3}
\hat{\Pi}_{\Phi}^{(N)}|\Phi^{(N)}\rangle=|\Phi^{(N)}\rangle\quad
\mbox{and}\quad
\hat{\Pi}_{\Psi}^{(N)}|\Psi^{(N)}\rangle=|\Psi^{(N)}\rangle.
\end{equation}
In particular, the superposition among  $|0\rangle$ and $|1\rangle$ in each subspace can be preserved through the measurement whereas the individual projection measurements make them collapsed. In other words, in the collective strategy, ``quantumness" can fully remain. The quantumness in the state can be used as resources for other quantum information processing. Because of the particular advantage, the collective measurement must be significant  not only from theoretical but also from engineering point of view. Unfortunately, however, implementation of the collective measurement as a physical system is not obvious. In this article, we propose an idea to use of quantum annealing mechanism to implement a collective projection measurement.

Before describing our idea, let us briefly review general idea of the quantum annealing. The goal of the quantum annealing is to dynamically obtain a ground state $|G_P\rangle$ of a given Hamiltonian $\hat{H}_P$. Considering a time dependent Hamiltonian $\hat{H}(t)$ that holds
\begin{equation}\label{eq:condition}
\hat{H}(0)=\hat{V},\quad
\hat{H}(T)=\hat{H}_P
\end{equation}
where $T>0$ is a final time of the annealing process and $\hat{V}$ is a driving Hamiltonian that is non-commutative with $\hat{H}_P$, one can show that the state evolved by Schr\"{o}dinger dynamics
\begin{equation}\label{eq:annealing process}
\frac{d}{dt}|\phi_t\rangle=-i\hat{H}(t)|\phi_t\rangle
\end{equation}
can approximately follow a ground state of the instantaneous Hamiltonian $\hat{H}(t)$ as $|\phi_t\rangle\simeq|G(t)\rangle$ when the initial state $|\phi_{t=0}\rangle$ is appropriately chosen to be the ground state of $\hat{V}$, and when $t$-dependence of $\hat{H}(t)$ satisfies the so-called adiabatic condition\cite{PhysRevE.58.5355}:
\begin{equation}\label{eq:condition7}
|\langle E_1(t) |\frac{d}{dt}\hat{H}(t) | G(t) \rangle| \ll 1
\end{equation}
where $|E_1(t)\rangle$ is the first excited state of the instantaneous Hamiltonian.
As the consequence of the above mechanism, $|\phi_t\rangle$ safely converges to $|G_P\rangle$.
When we employ a particular form of the time dependent Hamiltonian as
\begin{equation}\label{eq:annealing Hamiltonian}
\hat{H}(t)=\frac{t}{T}\hat{H}_P+\left(1-\frac{t}{T}\right) \hat{V}
\end{equation}
with large $T$, the conditions (\ref{eq:condition}) and (\ref{eq:condition7}) can be automatically satisfied. For instance, putting 
\begin{equation}\label{exampled Hamiltonian}
\hat{H}_P=h\Big(|1\rangle\langle 1|-|0\rangle\langle 0|\Big),\quad
\hat{V}=-\gamma\Big(|0\rangle\langle 1|+|1\rangle\langle 0|\Big)
\end{equation}
as the simplest example, we get
\begin{equation}\label{approximated solution}
|\phi_t\rangle\simeq \frac{-f(t)}{\sqrt{\gamma^2 (t-T)^2+f(t)^2}}|1\rangle -\frac{\gamma(t-T)}{\sqrt{\gamma^2 (t-T)^2+f(t)^2}} |0\rangle
\end{equation} 
with
\begin{equation}\label{approximated solution2}
f(t)=- ht+\sqrt{(ht)^2+\gamma^2 (t-T)^2}
\end{equation}
when we appropriately make a choice of $T$. With (\ref{approximated solution}) and (\ref{approximated solution2}), one can directly check the convergence to the ground state of $\hat{H}_P$ in (\ref{exampled Hamiltonian}), i.e.,
\begin{equation}\label{eq:convergence}
|\phi_t\rangle\rightarrow\left\{
\begin{array}{cc}
|0\rangle & \mbox{for}~h>0 \\
|1\rangle & \mbox{for}~h<0
\end{array},\right.
~\mbox{as}~t\rightarrow T.
\end{equation}

Our idea is to use the above annealing mechanism to drive the pointer state \cite{1981PhRvD..24.1516Z} associated with the measurement process. Depending on the given state ($|\Phi^{(N)}\rangle$ or $|\Psi^{(N)}\rangle$), effective $\hat{H}_P$ is adaptively introduced for the annealing process that the pointer state undergoes. 

To concretize the idea, we introduce the following Hamiltonian in Hilbert space ${\mathcal H}^{\otimes N}\otimes{\mathcal K}$ where ${\mathcal K}$ is a Hilbert space for the degree of freedom of the pointer state:
\begin{eqnarray}\label{measurement Hamiltonian}
\hat{H}_M(t)&=&\frac{t}{T}\sum_{j=1}^{N}\frac{2h_{j}(t)}{\epsilon}\bigotimes_{j'\not=j}^N \hat{I}_{j'}\otimes \Big(|1\rangle\langle 1|_j \otimes |1\rangle\langle 1|_{\mathcal K}+|0\rangle\langle 0|_j \otimes |0\rangle\langle 0|_{\mathcal K}\Big)\nonumber\\
&&\hspace*{2cm}-\gamma \left(1-\frac{t}{T}\right)\bigotimes_{j=1}^N\hat{I}_j\otimes \Big(|0\rangle\langle 1|_{\mathcal K}+|1\rangle\langle 0|_{\mathcal K}\Big)
\end{eqnarray}
where the operators suffixed by $j$ acts on the state in the $j$-th subspace in ${\mathcal H}^{\otimes N}$, and $h_{j}(t)$ is a time dependent coupling between the $j$-th state and the pointer state such as
\begin{equation}\label{step functions}
h_{j}(t)=\left\{
\begin{array}{lll}
0&\mbox{for}&t <\frac{T}{N}(j-1)\\
h(>0)&\mbox{for}&\frac{T}{N}(j-1)\le t < \frac{T}{N}j \\
0&\mbox{for}&\frac{T}{N}j \le t.\\
\end{array}
\right.
\end{equation}
In order to get intuitive implications of the above Hamiltonian, let us see $\langle \Phi^{(N)}|\hat{H}_M(t)|\Phi^{(N)}\rangle$ and  $\langle \Psi^{(N)}|\hat{H}_M(t)|\Psi^{(N)}\rangle$. We find that
\begin{equation}\label{effective Hamiltonian1}
\langle \Phi^{(N)}|\hat{H}_M(t)|\Phi^{(N)}\rangle=-\frac{t}{T}h\Big(|1\rangle\langle 1|_{\mathcal K}-|0\rangle\langle 0|_{\mathcal K}\Big)-\gamma\left(1-\frac{t}{T}\right)\Big(|0\rangle\langle 1|_{\mathcal K}+|1\rangle\langle 0|_{\mathcal K}\Big)
\end{equation}
and
\begin{equation}\label{effective Hamiltonian2}
\langle \Psi^{(N)}|\hat{H}_M(t)|\Psi^{(N)}\rangle=+\frac{t}{T}h\Big(|1\rangle\langle 1|_{\mathcal K}-|0\rangle\langle 0|_{\mathcal K}\Big)-\gamma\left(1-\frac{t}{T}\right)\Big(|0\rangle\langle 1|_{\mathcal K}+|1\rangle\langle 0|_{\mathcal K}\Big).
\end{equation}
Compare (\ref{effective Hamiltonian1}) and (\ref{effective Hamiltonian2}) with Eqs. (\ref{eq:annealing Hamiltonian}), (\ref{exampled Hamiltonian}) and (\ref{eq:convergence}). The comparison suggests that, {\it if we can ignore the time evolution of the state in ${\mathcal H}^{\otimes N}$, the Hamiltonian in (\ref{measurement Hamiltonian})  acts as an effective annealing Hamiltonian for the pointer state in ${\mathcal K}$ which deterministically drives the state to $|1\rangle_{\mathcal K}$ or $|0\rangle_{\mathcal K}$ depending on the given state in ${\mathcal H}^{\otimes N}$.} In general, the time evolution of the state in ${\mathcal H}^{\otimes N}$, which can be interpreted as a back reaction caused by the measurement, cannot be simply ignored. Taking large $N$, however, we can bound the duration of the interaction between the $j$-th state and the pointer state. We are able to make the back reaction to diminish. Moreover, there is a chance to make 
\begin{equation}
N \times \mbox{back reaction to each state}
\end{equation}
to be negligible. As is numerically demonstrated  below, we can achieve the above situation indeed. 

Let us numerically examine our idea as follows: We numerically solve  Schr\"{o}dinger equation
\begin{equation}\label{schrodinger equation for measurement}
\frac{d}{dt}|\Xi(t)\rangle=-i \hat{H}_M(t) |\Xi(t)\rangle,\quad |\Xi(t)\rangle\in {\mathcal H}^{\otimes N}\otimes{\mathcal K}
\end{equation}
with the two initial states:
\begin{equation}
\mbox{Case $\Phi$:}~~  |\Xi(0)\rangle=|\Phi^{(N)}\rangle\otimes\frac{1}{\sqrt{2}}\Big(|0\rangle+|1\rangle\Big),\quad
\mbox{Case $\Psi$:}~~  |\Xi(0)\rangle=|\Psi^{(N)}\rangle\otimes\frac{1}{\sqrt{2}}\Big(|0\rangle+|1\rangle\Big)
\end{equation}
Corresponding to each case, let us put index on solution $|\Xi(t)\rangle$ as $|\Xi^{\Phi^{(N)}}(t)\rangle$ or $|\Xi^{\Psi^{(N)}}(t)\rangle$. From the solution, dynamics of the pointer state is obtained as
\begin{equation}\label{state of pointer state}
\left\{
\begin{array}{ccc}
\hat{\rho}_{\Phi^{(N)}}(t)&=&{\rm tr}_{{\mathcal H}^{\otimes N}}\Big(|\Xi^{\Phi^{(N)}}(t)\rangle\langle\Xi^{\Phi^{(N)}}(t)|\Big)\\
\hat{\rho}_{\Psi^{(N)}}(t)&=&{\rm tr}_{{\mathcal H}^{\otimes N}}\Big(|\Xi^{\Psi^{(N)}}(t)\rangle\langle\Xi^{\Psi^{(N)}}(t)|\Big)
\end{array}\right.
\end{equation}
where ${\rm tr}_{{\mathcal H}^{\otimes N}}$ denotes the partial trace operation over Hilbert space ${\mathcal H}^{\otimes N}$. Similarly, the state in ${\mathcal H}^{\otimes N}$ after the annealing process is obtained as
\begin{equation}
\left\{
\begin{array}{ccc}
\hat{\mu}_{\Phi^{(N)}}(T)&=&{\rm tr}_{\mathcal K}\Big(|\Xi^{\Phi^{(N)}}(t)\rangle\langle\Xi^{\Phi^{(N)}}(t)|\Big)\\
\hat{\mu}_{\Psi^{(N)}}(T)&=&{\rm tr}_{\mathcal K}\Big(|\Xi^{\Psi^{(N)}}(t)\rangle\langle\Xi^{\Psi^{(N)}}(t)|\Big)
\end{array}\right.
\end{equation}
where ${\rm tr}_{\mathcal K}$ denotes the partial trace operation over Hilbert space ${\mathcal K}$. Corresponding to the properties of the projection measurement described in (\ref{property 1}), (\ref{property 2}) and (\ref{property 3}), if
we have 
\begin{equation}\label{property 4}
\langle 1|\hat{\rho}_{\Phi^{(N)}}(T) |1\rangle_{\mathcal K}=\langle 0|\hat{\rho}_{\Psi^{(N)}}(T) |0\rangle_{\mathcal K}=1,
\end{equation}
\begin{equation}\label{property 5}
\langle 0|\hat{\rho}_{\Phi^{(N)}}(T) |0\rangle_{\mathcal K}=\langle 1|\hat{\rho}_{\Psi^{(N)}}(T) |1\rangle_{\mathcal K}=0,
\end{equation}
and
\begin{equation}\label{property 6}
\langle \Phi^{(N)}|\hat{\mu}_{\Phi^{(N)}}(T)|\Phi^{(N)}\rangle=
\langle \Psi^{(N)}|\hat{\mu}_{\Psi^{(N)}}(T)|\Psi^{(N)}\rangle=1,
\end{equation}
we can safely claim that our process can be regarded as the collective projection measurement. In the following numerical analysis, $h=\gamma=\frac{1}{2}$ and $T=10$ are commonly used.

Concerning the properties in (\ref{property 4}) and (\ref{property 5}), we numerically obtain results as shown in  Figs.1, 2 and 3. The time evolution of the pointer state given in (\ref{state of pointer state}) is shown in Fig.1. We can find that, when $N$ is sufficiently large, the pointer state is driven to $|1\rangle_{\mathcal K}$ (which indicates the given state is $|\Phi^{(N)}\rangle$) as is expected. In Fig.2, convergence of $\langle 1|\hat{\rho}_{\Phi^{(N)}}(t)|1\rangle_{\mathcal K}$ to $|\langle 1|\phi_t\rangle|^2$ with large $N$ can be found where $|\phi_t\rangle$ is the solution of the original annealing process described in (\ref{eq:annealing process}), (\ref{eq:annealing Hamiltonian}) and (\ref{exampled Hamiltonian}).  Notice that  one can also find a scaling law with a scaling parameter 
\begin{equation}\label{scaling law}
\lambda:=N\epsilon^2
\end{equation}
in Fig.2. For instance, the third line from the top in Fig.2(a) with $(\epsilon,N)=(2^{-6}, 2^{12})$ is almost identical to the last line in Fig.2(b) with $(\epsilon,N)=(2^{-8}, 2^{16})$. This scaling law can be even more clearly observed in Fig.3, in which minimum $N$ required for 
\begin{equation}
\langle 1|\hat{\rho}_{\Phi^{(N)}}(T)|1\rangle_{\mathcal K}\ge P_{success}
\end{equation}
 (with $P_{success}=0.9$) is plotted with respect to $\epsilon$. The scaling law fits them quite well, and larger $P_{success}$ requires larger $\lambda$ to be fitted. The relation between $P_{success}$ and $\lambda$ is a direct reflection of the condition in (\ref{f-condition}) that is nothing but a fundamental limit of (in)distinguishability of (non)orthogonal states in quantum theory.
\begin{figure}[H]
\centering
\begin{subfigure}[t]{0.3\textwidth}
\centering
\includegraphics[height=35mm,width=45mm]{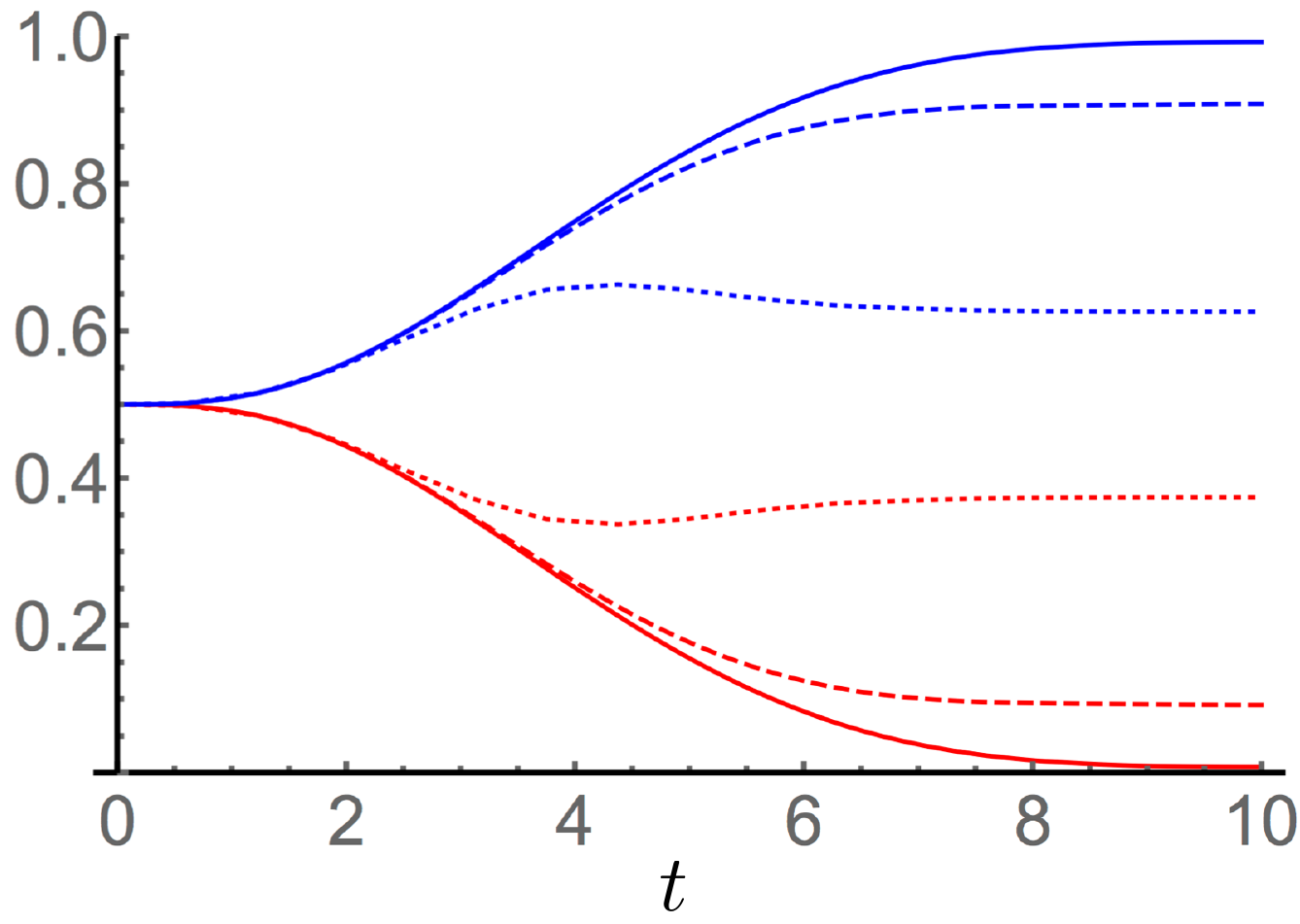}
\caption{$\epsilon=1/4$}
\end{subfigure}
~~~~~~~~~~~~~~~~~~~~
\begin{subfigure}[t]{0.3\textwidth}
\centering
\includegraphics[height=35mm,width=45mm]{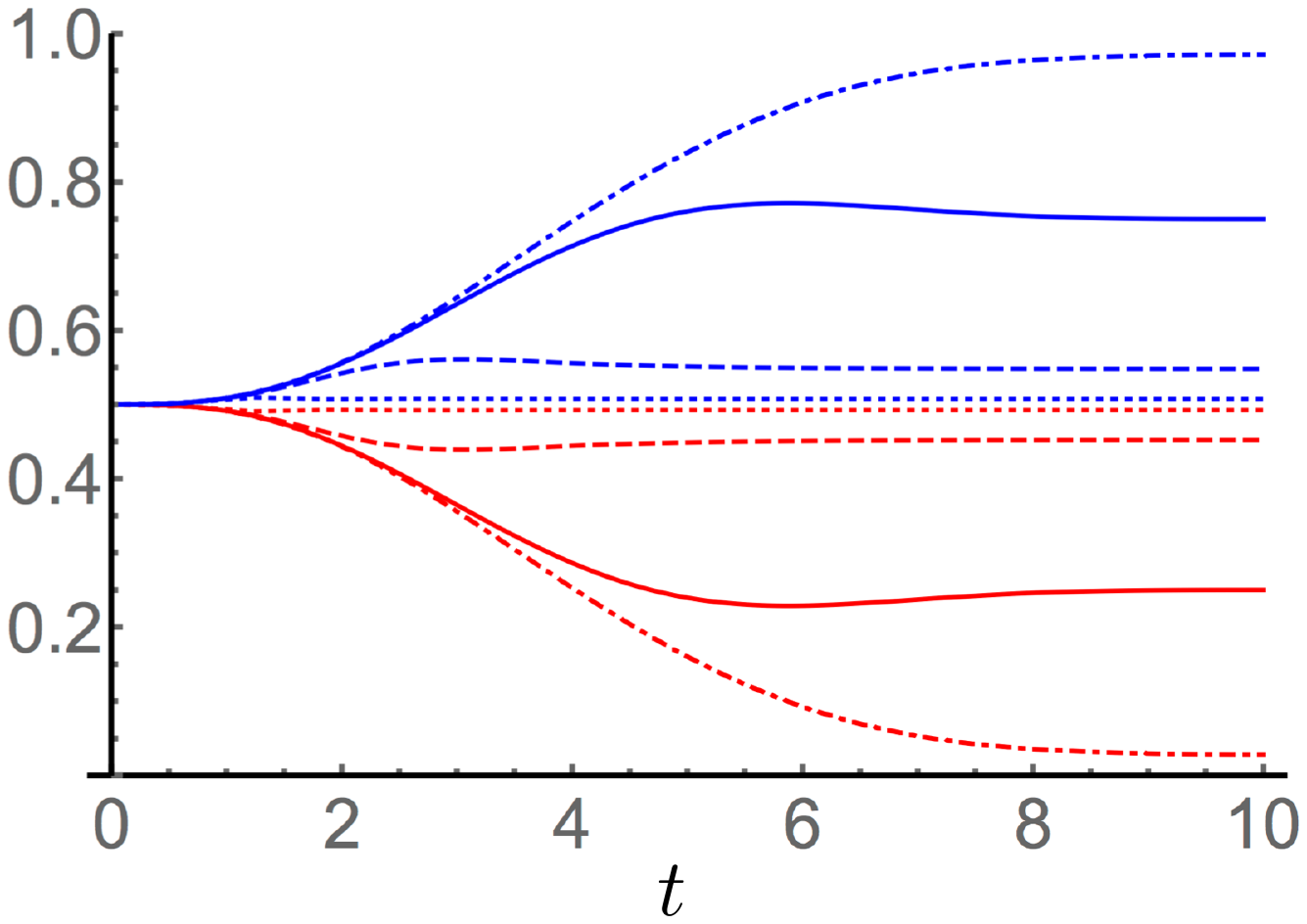}
\caption{$\epsilon=1/32$}
\end{subfigure}
\caption{{\bf $\epsilon$-dependence of pointer state $\hat{\rho}_{\Phi^{(N)}}(t)$ with $h=\gamma=1/2$ and $T=10$.} Diagonal elements $\langle 1|\hat{\rho}_{\Phi^{(N)}}(t)|1\rangle_{\mathcal K}$ and $\langle 0|\hat{\rho}_{\Phi^{(N)}}(t)|0\rangle_{\mathcal K}$ are plotted in blue and in red respectively. In (a) and (b), dotted, dashed, and solid lines correspond to $N=16$, ~$256,$
and $4096$ respectively. In addition, the case of $N=65536$ is shown in dot-dashed line in (b).}
\end{figure}
\begin{figure}[H]
\centering
\begin{subfigure}[t]{0.3\textwidth}
\centering
\includegraphics[height=35mm,width=45mm]{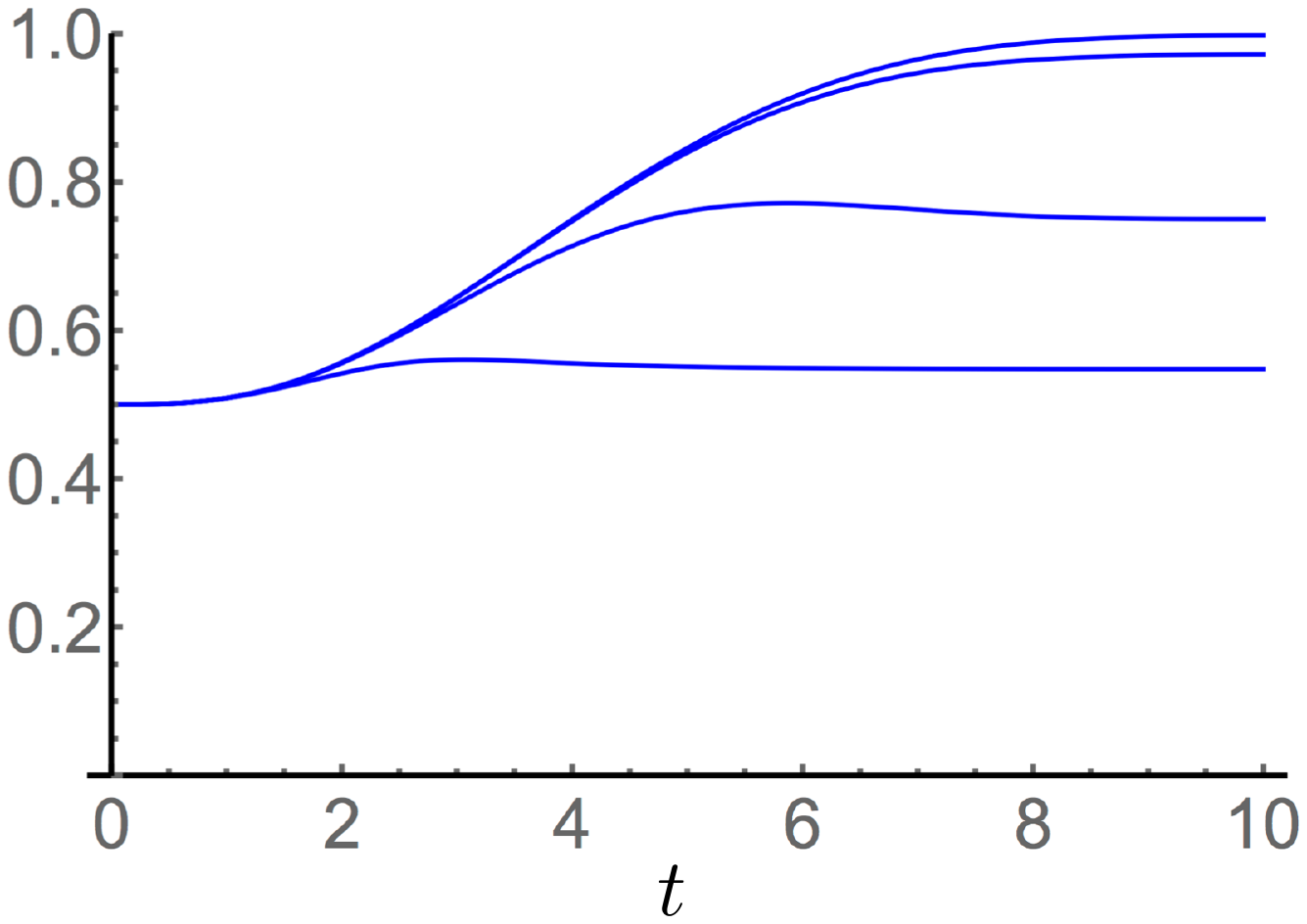}
\caption{$N=4096$}
\end{subfigure}
~
\begin{subfigure}[t]{0.3\textwidth}
\centering
\includegraphics[height=35mm,width=45mm]{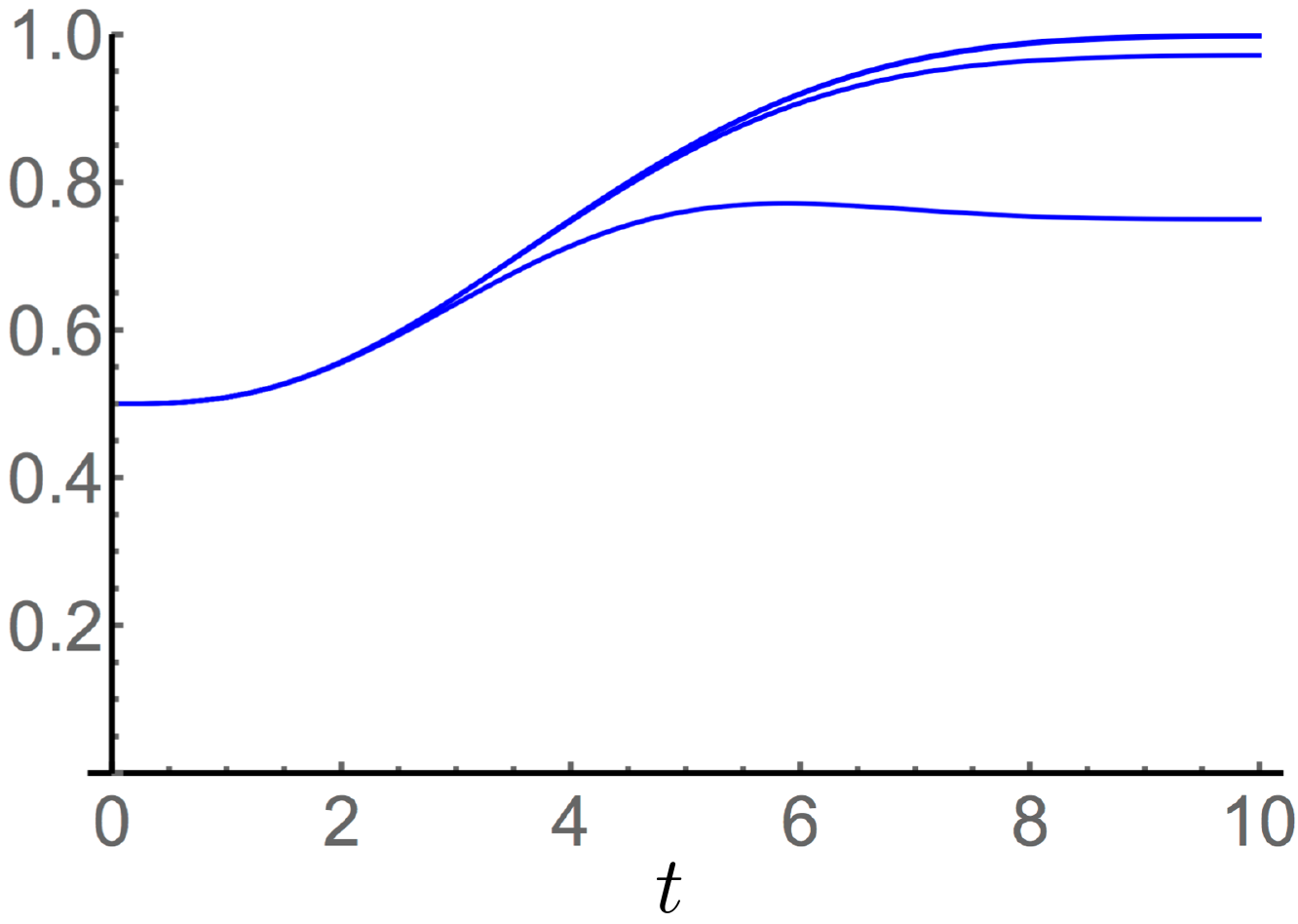}
\caption{$N=65536$}
\end{subfigure}
~
\begin{subfigure}[t]{0.3\textwidth}
\centering
\includegraphics[height=35mm,width=45mm]{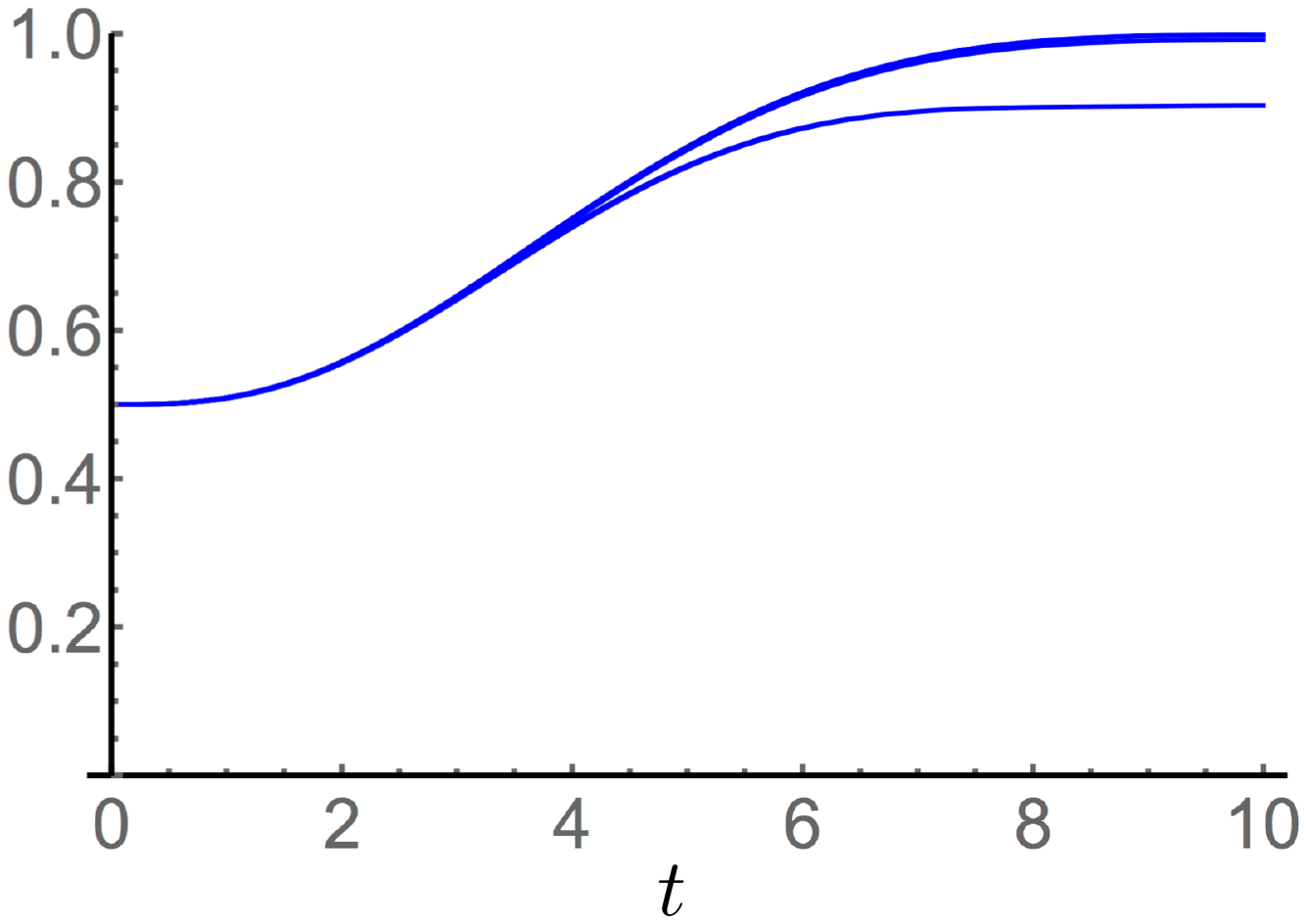}
\caption{$N=262144$}
\end{subfigure}
\caption{{\bf $N$-dependence in pointer state $\hat{\rho}_{\Phi^{(N)}}(t)$}. Diagonal elements $\langle 1|\hat{\rho}_{\Phi^{(N)}}(t)|1\rangle_{\mathcal K}$ with $\epsilon=1/2, 1/8, 1/32$, and $1/128$ are plotted from top to bottom. In (b) and (c), the top two and three lines are almost overlapped respectively.}
\end{figure}
\begin{figure}[H]
\centering
\includegraphics[height=49mm,width=63mm]{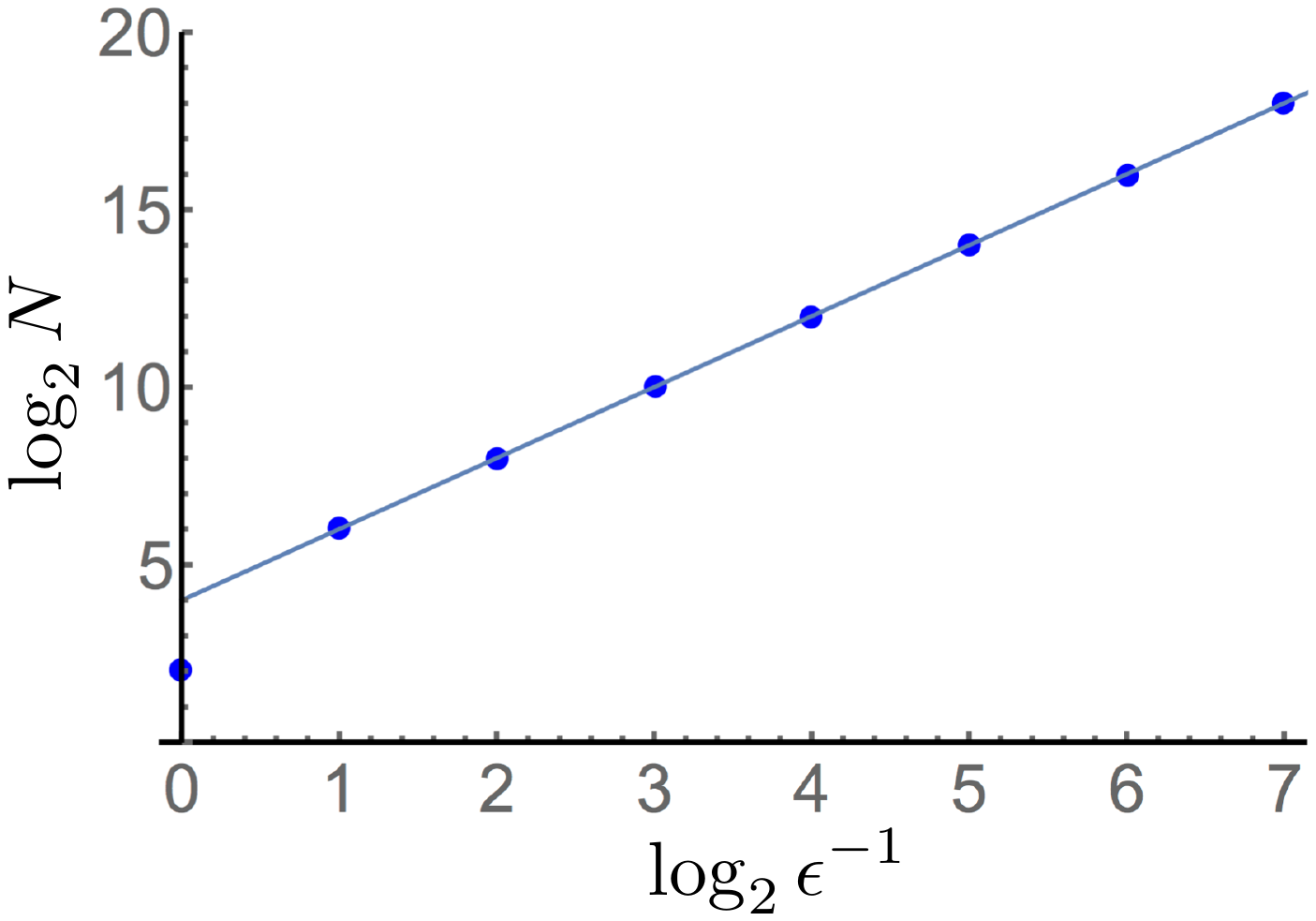}
\caption{{\bf $\epsilon$-dependence in $N$ for success measurement}. Minimum $N$ required for $\langle 1|\hat{\rho}_{\Phi^{(N)}}(T)|1\rangle_{\mathcal K}\ge 0.9$ is plotted with respect to $\epsilon$. Solid line shows (\ref{scaling law}) with $\lambda=16.$}
\end{figure}

Let us also check the projection property described in (\ref{property 6}). In Fig.4, the fidelity between $|\Phi^{(N)}\rangle$ 
(state before being measured) and $\hat{\mu}_{\Phi^{(N)}}(T)$ (state after being measured) is numerically shown with respect to $N$. When $N$ is not large enough, the process cannot preserve the state except for the cases of $\epsilon=0$ or $1$. (The reason that the state with $\epsilon=0$ or $1$ is always preserved is that the state behaves as the eigenstate of  $\hat{H}_{M}(t)$ in its subspace ${\mathcal H}^{\otimes N}$ independently from $N$. ) The fidelity suddenly tends to $1$ as $N$ becomes larger than a threshold that depends on $\epsilon$. Combining the result from Figs.1, 2, 3 and 4, we can safely say that the dynamics proposed by (\ref{measurement Hamiltonian}), (\ref{step functions}) and (\ref{schrodinger equation for measurement}) works as the projection measurement in the collective space when $N$ is large enough. Figure 5 shows the $\epsilon$-dependence of $N$ which is required to achieve
\begin{equation}
\langle \Phi^{(N)}|\hat{\mu}_{\Phi^{(N)}}(T)|\Phi^{(N)}\rangle_{{\mathcal H}^{\otimes N}} \ge F_{Fidelity}
\end{equation}
(with $F_{Fidelity}=0.9$). We find the  scaling law in (\ref{scaling law}) here too. We would like to underline that our realization of the projection measurement is ``tight" in the sense that the success of the measurement rely only on the fundamental condition in (\ref{f-condition}). 
\begin{figure}[H]
\centering
\includegraphics[height=49mm,width=63mm]{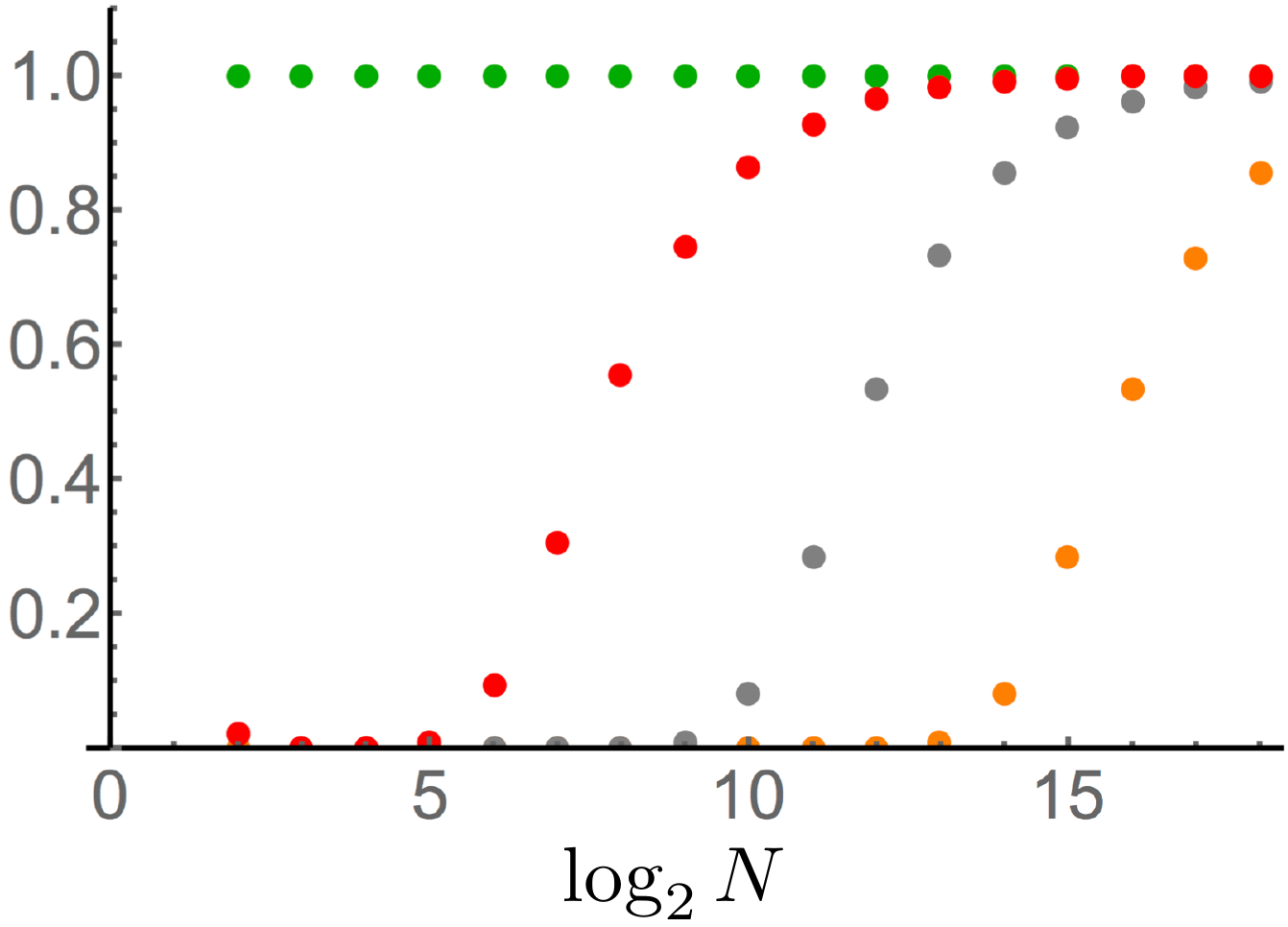}
\caption{{\bf Fidelity between before and after}. $N$-dependence of  $\langle \Phi^{(N)}|\hat{\mu}_{\Phi^{(N)}}(T)|\Phi^{(N)}\rangle_{{\mathcal H}^{\otimes N}}$ with $\epsilon=1, ~1/4, ~1/16$ and $1/64$ are plotted in green, red, gray, and orange respectively.}
\end{figure}

\begin{figure}[H]
\centering
\includegraphics[height=49mm,width=63mm]{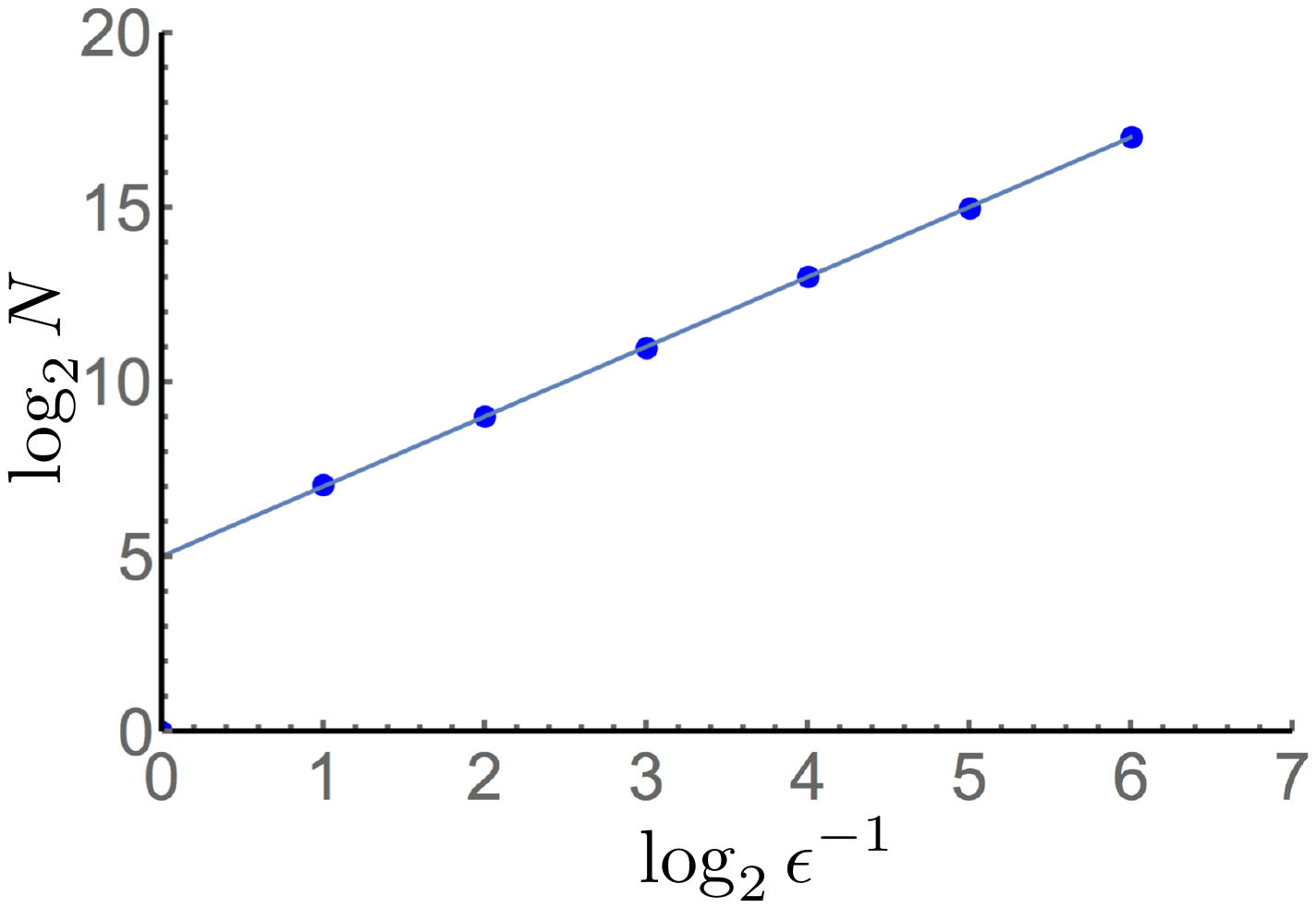}
\caption{{\bf $\epsilon$-dependence in $N$ for projection property}. Minimum $N$ required for $\langle \Phi^{(N)}|\hat{\mu}_{\Phi^{(N)}}(T)|\Phi^{(N)}\rangle_{{\mathcal H}^{\otimes N}}\ge 0.9$ is plotted with respect to $\epsilon$. Solid line shows (\ref{scaling law}) with $\lambda=32$.}
\end{figure}

Finally, let us sum up our approach in a context of the quantum annealing mechanism. What we realized by (\ref{measurement Hamiltonian}) is a quantum annealing process in the subspace ${\mathcal K}$ by giving its problem Hamiltonian depending on the quantum state $|\Phi^{(N)}\rangle$ or $|\Psi^{(N)}\rangle$. In other words, the parameters in the problem Hamiltonian is given not as classical variables but as quantum variables (states). By additionally introducing successive short interactions so that the back reaction to the quantum states giving the parameters can be controlled, we  invent a quantum mechanically parametrized quantum annealing process. Applying the particular problem of discrimination of two collective states , we find that the process by the quantum mechanically parametrized annealing arrives at projection measurement in the collective space when the parametrizing quantum variables themselves are orthogonal (or distinguishable).  We believe that this can be a (small but interesting)  extension of the concept of the quantum annealing mechanism to make it applicable to broader applications than the conventional one. We are looking forward to introducing some applications besides the particular problem we discussed in this article.

\bibliographystyle{unsrt}
\bibliography{aqc2018}

\begin{thebibliography}{1}

\bibitem{PhysRevB.39.11828}
P.~Ray, B.~K. Chakrabarti, and Arunava Chakrabarti.
\newblock Sherrington-kirkpatrick model in a transverse field: Absence of
  replica symmetry breaking due to quantum fluctuations.
\newblock {\em Phys. Rev. B}, 39:11828--11832, Jun 1989.

\bibitem{PhysRevE.58.5355}
Tadashi Kadowaki and Hidetoshi Nishimori.
\newblock Quantum annealing in the transverse ising model.
\newblock {\em Phys. Rev. E}, 58:5355--5363, Nov 1998.

\bibitem{2000quant.ph..1106F}
E.~{Farhi}, J.~{Goldstone}, S.~{Gutmann}, and M.~{Sipser}.
\newblock {Quantum Computation by Adiabatic Evolution}.
\newblock {\em eprint arXiv:quant-ph/0001106}, January 2000.

\bibitem{2014McGeoch}
Catherine~C. McGeoch.
\newblock {\em Adiabatic Quantum Computation and Quantum Annealing: Theory and
  Practice}.
\newblock Synthesis Lectures on Quantum Computing. Morgan {\&} Claypool
  Publishers, 2014.

\bibitem{von2018mathematical}
J.~von Neumann, R.T. Beyer, and N.A. Wheeler.
\newblock {\em Mathematical Foundations of Quantum Mechanics: New Edition}.
\newblock Princeton University Press, 2018.

\bibitem{1981PhRvD..24.1516Z}
W.~H. {Zurek}.
\newblock {Pointer basis of quantum apparatus: Into what mixture does the wave
  packet collapse?}
\newblock {\em Phys. Rev. D}, 24:1516--1525, September 1981.

\end{thebibliography}

\end{document}